\documentclass[10pt,final]{IEEEtran}

\usepackage[utf8]{inputenc}
\usepackage{graphicx}
\usepackage{stfloats}
\usepackage{amssymb}
\usepackage[cmex10]{amsmath}
\usepackage{color}
\usepackage{pifont}
\usepackage{array}
\usepackage{hyperref}
\usepackage{cite}
\usepackage{url}
\usepackage{bbm}
\usepackage{relsize}
\usepackage[ruled,vlined,titlenumbered]{algorithm2e_old2}

\usepackage{ifthen}

\newcommand{\F}[1]{\ensuremath{\mathbb{F}_{#1}}}
\newcommand{\Fq}{\F{q}}
\newcommand{\Fx}[1]{\ensuremath{\F{#1}[x]}}
\newcommand{\Fxq}{\Fx{q}}

\newcommand{\refeq}[1]{(\ref{#1})}

\newtheorem{definition}{Definition}
\newtheorem{theorem}{Theorem}
\newtheorem{lemma}{Lemma}

\newtheorem{question}{Question}
\newtheorem{example}{Example}

\newcommand{\printalgo}[1]
{
\begin{center}
\scalebox{0.9}{
\begin{algorithm}[H]
 #1
\end{algorithm}
}
\end{center}
}

\newcommand{\mat}[2][\empty]{
  \ifthenelse{\equal{#1}{\empty}}
    {\ensuremath{\mathbf{#2}}}
    {\ensuremath{#2_{#1}}}
}

\newcommand{\vect}[2][\empty]{
  \ifthenelse{\equal{#1}{\empty}}
    {\ensuremath{\mathbf{#2}}}
    {\ensuremath{#2_{#1}}}
}

\renewcommand{\c}{\mathbf c}

\DeclareMathOperator{\defi}{def}
\newcommand{\defeq}{\overset{\defi}{=}}

\renewcommand{\bar}{\overline}
\renewcommand{\hat}{\widehat}

\newcommand{\CYC}[4]{\ensuremath{\mathcal{C}(#1; #2,#3,#4)}}
\newcommand{\defset}{\ensuremath{D_\mathcal{C}}}
\newcommand{\cw}{\ensuremath{(c_0 \ c_1 \ \dots \ c_{n-1})}}

\newlength{\defrowheight}
\newcommand{\muHT}{\ensuremath{d_0}}
\newcommand{\nuHT}{\ensuremath{\nu}}
\newcommand{\HTa}{\ensuremath{m_1}}
\newcommand{\HTb}{\ensuremath{m_2}}
\newcommand{\fracjump}{\ensuremath{z_1}}

\newcommand{\dBCH}{\ensuremath{d_{\text{BCH}}}}
\newcommand{\dHT}{\ensuremath{d_{\text{HT}}}}
\newcommand{\dBOSTON}{\ensuremath{d_{\text{B}}}}

\begin{document}

\title{Decoding Cyclic Codes up to a New\\ Bound on the Minimum Distance}

\IEEEoverridecommandlockouts

\author{{Alexander Zeh, Antonia Wachter-Zeh, and Sergey Bezzateev \thanks{The material in this contribution was presented in part at the IEEE International Symposium on Information Theory (ISIT 2011) in St. Petersburg, Russia~\cite{ZehWachterBezzateev_EfficientDecodingOfSomeClasses_2011}. This work has been supported by DFG, Germany, under grants BO~867/22-1 and BO~867/21-1.

A. Zeh is with the Institute of Communications Engineering, University of Ulm, D-89081 Ulm, Germany and
with Research Center INRIA Saclay - \^{I}le-de-France, École Polytechnique ParisTech, 91128 Palaiseau Cedex, France
(e-mail: alexander.zeh@uni-ulm.de).

A. Wachter-Zeh is with the Institute of Communications Engineering, University of Ulm, D-89081 Ulm, Germany
and Institut de Recherche Mathémathique de Rennes (IRMAR), Université de Rennes 1, 35042 Rennes Cedex, France
(e-mail: antonia.wachter@uni-ulm.de).

S. Bezzateev is with the Department of Information Systems and Security, Saint Petersburg State University of Airspace Instrumentation, Saint-Petersburg, Russia (e-mail: bsv@aanet.ru).
}}}

\maketitle

\begin{abstract}
A new lower bound on the minimum distance of $q$-ary cyclic codes is proposed.
This bound improves upon the Bose--Chaudhuri--Hocquenghem (BCH) bound
and, for some codes, upon the Hartmann--Tzeng (HT) bound. Several Boston
bounds are special cases of our bound. For some classes of codes the bound on the minimum distance is refined. Furthermore, a quadratic-time decoding algorithm up to this new bound is developed.
The determination of the error locations is based on the Euclidean Algorithm and a modified
Chien search. The error evaluation is done by solving a generalization of Forney's formula.
\end{abstract}


\section{Introduction} \label{sec_intro}
\IEEEPARstart{S}{everal} bounds on the minimum distance of
cyclic codes are defined by a subset of the defining set of the
code. The Bose--Chaudhuri--Hocquenghem (BCH) bound
\cite{Hocquenghem_1959, Bose_RayChaudhuri_1960} considers \emph{one} set of \emph{consecutive} elements of the defining
set. A first extension of this bound was formulated by Hartmann
and Tzeng (HT) \cite{Hartmann_DecodingBeyondBCHbound_1972,
Hartmann_GeneralizationsofBCHbound_1972,
Hartmann_DecodingBeyondBCHBound_1974,
Hartmann_SomeResultMinimumDistance_1972}, where \emph{several}
sets of \emph{consecutive} elements are used to increase the
lower bound on the minimum distance. The Roos bound
\cite{Roos_GeneralBCHBound_1982,Roos_BoundforCyclicCodes_1983}
generalizes this idea by exploiting \emph{several} sets
of \emph{nonconsecutive} elements in the defining set.
The contributions of van Lint and Wilson~\cite{vanLint_OnTheMinimumDistance_1986},
Duursma and K\"otter~\cite{DuursmaKoetter_ErrorLocatingPairs_1994} and
Duursma and Pellikann~\cite{Duursma_ASymmetricRoos_2006} are further
generalizations.
Other approaches include the Boston bounds~\cite{Boston_CyclicCodesAlgebraicGeometry_2001} and the bound by Betti and Sala~\cite{BettiSala_NewLowerBound_2006}.

Although these improved bounds show that for many codes the actual distance is
higher than the BCH bound, there is no general decoding algorithm up to any of these bounds.
Hartmann and Tzeng~\cite{Hartmann_DecodingBeyondBCHbound_1972, Hartmann_DecodingBeyondBCHBound_1974} proposed two variants of an
iterative decoding algorithm up to the HT bound. However, these algorithms require the calculation of
missing syndromes and the solution of non-linear equations.
An approach for decoding all binary cyclic codes up to their \emph{actual} minimum distance of length less than 63
was given by Feng and Tzeng \cite{Feng_Tzeng_1994}. They use a
generalized syndrome matrix and fit the known syndrome coefficients manually for
each code into the structure of the matrix.

This contribution provides a new lower bound on the minimum distance of $q$-ary cyclic codes 
based on a connection of the code with rational functions.
This approach originates from decoding Goppa
codes~\cite{Goppa_RationReprCodes_1971 ,Goppa_NewClassCodes_1970,Bezzateev_OneGenGoppaCodes_1997,Shekhunova_LGCodes_1981}.
We match the roots of a $q$-ary cyclic code to nonzeros of the power series expansion of a
rational function. This allows to formulate a new lower bound on the minimum distance
of cyclic codes. We identify some classes of cyclic codes and refine the bound on their distance.
A wide class of codes, which is covered by our
approach, is the class of reversible codes \cite{Massey_ReversibleCodes_1964}.
Our new lower bound is better than the BCH bound and for most codes also better than
the HT bound. Moreover, it can be seen as a generalization of some Boston~\cite{Boston_CyclicCodesAlgebraicGeometry_2001} bounds.
We give tables for binary and ternary cyclic codes, where we count the number of
cyclic codes for which our bound is better than the BCH bound.

As a second part, we give an efficient decoding algorithm
up to our new bound.
This decoding algorithm is based on a generalized key equation, a modified
Chien search and a generalized Forney's formula~\cite{Forney_OndecodingBCHcodes_1965} for the error evaluation.
The time complexity of the whole decoding procedure is quadratic with the length of the
cyclic code.

This contribution is structured as follows.
Section~\ref{sec_preliminaries} gives some basic definitions and recapitulates known bounds on the minimum
distance of cyclic codes. We show how the
BCH bound can be represented by a simple rational function.
In Section~\ref{sec_principle}, we explain how we associate a rational function to a cyclic code and we prove our new lower bound on the minimum distance.
Section~\ref{sec_classes} provides several identified classes and we refine
the lower bound of these codes. We compare our new lower bound on the minimum distance with the
BCH and the HT bound.
In Section~\ref{sec_boston}, we show how several Boston bounds are generalized by
our principle.
The decoding algorithm is given in Section~\ref{sec_decoding}. Therefore, a generalized key
equation is derived and the decoding radius is proved.
Section~\ref{sec_conclusion} concludes this contribution.

%
%
%
%

\section{Preliminaries} \label{sec_preliminaries}
\subsection{Q-Ary Cyclic Codes and Rational Functions}
Let $q$ be a power of a prime, let $\Fq$ denote the finite field of order $q$ and let $\Fxq$ denote the set of all univariate polynomials with coefficients in $\Fq$ and the indeterminate $x$.
A $q$-ary cyclic code of length $n$, dimension $k$ and minimum distance $d$ is
denoted by \CYC{\Fq}{n}{k}{d}. A codeword of \CYC{\Fq}{n}{k}{d} is a multiple of its generator polynomial $g(x)$
with roots in \F{q^s}, where $n \mid (q^s-1)$.  Let $\alpha$ be an $n$th root of unity of $\F{q^s}$.
A cyclotomic coset $M_r$ is given by:
\begin{equation} \label{eq_cyclotomiccoset}
 M_r = \lbrace rq^j \bmod n, \; \forall j = 0,1,\dots,n_r-1 \rbrace,
\end{equation}
where $n_r$ is the smallest integer such that $rq^{n_r} \equiv
r \mod n$. It is well-known that the minimal polynomial $M_r(x) \in \Fxq$ of the element $\alpha^r$
is given by
\begin{equation} \label{eq_minpoly}
 M_r(x) = \prod_{i \in M_r} (x-\alpha^i).
\end{equation}
The defining set $D_{\mathcal{C}}$ of a $q$-ary cyclic code
\CYC{\Fq}{n}{k}{d} is the set containing the indices of the zeros of the generator
polynomial $g(x)$ and can be partitioned into $w$ cyclotomic
cosets:
\begin{equation} \label{eq_definingset}
\begin{split}
D_{\mathcal{C}} & \defeq   \{i : \, g(\alpha^i)=0 \} = M_{r_1} \cup M_{r_2} \cup \dots \cup M_{r_{w}}.
\end{split}
\end{equation}
Hence, the generator polynomial $g(x) \in \Fxq$ of degree $n-k$ of \CYC{\Fq}{n}{k}{d} is
\begin{equation}
 g(x) = \prod_{i=1}^{w} M_{r_i}(x).
\end{equation}
The following lemma states the cardinality of all cyclotomic cosets $M_r$,
if $r$ is co-prime to the length $n$. We use it later to determine the rate of
some classes of cyclic codes.
\begin{lemma}[Cardinality]\label{lem:cardinality_general}
Let $s$ be the smallest integer such that the length $n$
divides $(q^{s}-1)$, then the cardinality of the cyclotomic
coset $M_r$ is $|M_r|=s$ if $\gcd(n,r)=1$.
\end{lemma}
\begin{IEEEproof}
The cyclotomic coset $M_r$ has cardinality $|M_r|=j$ if and only if $j$ is the smallest integer such that
\begin{equation*}
r \cdot q^j \equiv r \mod n \quad \Longleftrightarrow \quad r \cdot (q^j-1) \equiv 0 \mod n.
\end{equation*}
Since $\gcd(n,r)=1$, this is equivalent to $n \mid (q^j-1)$.
Since $s$ is the smallest integer such that the length $n$
divides $(q^{s}-1)$, $j=s$ and hence, $|M_r|=s$.
\end{IEEEproof}

Let us state some preliminaries on rational functions.
\begin{definition}[Period of a Power Series] \label{def_periodrational}
Let a formal power series $a(x) = \sum_{j=0}^{\infty} a_j x^j$
with $a_j \in \F{q}$ be given. The period $p(a(x))$ of the infinite
sequence $a(x)$ is the smallest $p$, such that
\begin{equation*}
 a(x)= \frac{\sum_{j=0}^{p-1} a_j x ^j }{-x^{p}+1}
\end{equation*}
holds.
\end{definition}
Throughout this paper we use the power series expansion of the
fraction of two polynomials $h(x)$ and $f(x)$ in $\Fx{q}$ with
\begin{equation} \label{eq_degree_numeratordenominator}
 v \defeq \deg h(x) < u \defeq \deg f(x).
\end{equation}
We require that:
\begin{enumerate}
\item $\deg \gcd (h(x),f(x)) = 0$ and
\item $\deg \gcd (f(x\alpha^i),f(x\alpha^j)) = 0, \quad \forall i \neq j, \alpha^i,\alpha^j \in \F{q^s}$
\end{enumerate}
to prove our main theorem on the minimum distance.

The following lemma establishes a connection between the length $n$ of the code and the period
of the power series $h(x)/f(x)$, such that 2) holds.
\begin{lemma}[Code Length, Period of a Power Series] \label{lem_coprime}
Let $\alpha$ be an $n$th root of unity of
$\F{q^s}$, where $n\mid(q^s-1)$. Let $h(x),f(x) \in \Fxq$ with $\deg \gcd
(h(x),f(x)) = 0$ and degree as in~\refeq{eq_degree_numeratordenominator} be given. The formal power series is
$h(x)/f(x) \defeq \sum_{j=0}^{\infty} a_j x^j$ over $\F{q}$ with
period $p(h(x)/f(x)) = p$. If  the period $p$ and
$n$ are co-prime then
\begin{equation*}
\deg \gcd(f(x\alpha^i), f(x\alpha^j)) = 0, \ \forall i \neq j.
\end{equation*}
\end{lemma}
\begin{IEEEproof}
From Definition~\ref{def_periodrational}, we have
\begin{equation*}
h(x) (-x^p+1)=f(x)(a_0 + a_1 x + \ldots+ a_{p-1}  x^{p-1}),
\end{equation*}
and from $\deg \gcd(f(x), h(x))=0$, it follows that $-x^{p}+1
\equiv 0 \mod f(x)$. Hence, for two different polynomials
$f(x\alpha^i)$ and $f(x\alpha^j)$, for any $i\neq j,\;
i,j=0,\ldots,n-1$:
\begin{align} \label{lemma_gcd_fi_fj}
x^{p}\alpha^{ip}-1 & \equiv 0 \mod f(x\alpha^{i}) \quad \text{and} \nonumber \\
x^{p}\alpha^{jp}-1 & \equiv 0 \mod f(x\alpha^{j}).
\end{align}
Assume there is some
element $\beta \in \F{q^{us}}\setminus \{0\}$,
such that
$$
f(\beta\alpha^{i})=f(\beta\alpha^{j})=0, $$
$$ \text{ i.e.,} \;
\gcd(f(x\alpha^{i}), f(x\alpha^j)) \equiv 0 \mod (x-\beta).
$$
Equation~\eqref{lemma_gcd_fi_fj} gives the following:
$$
\beta^{p}\alpha^{ip}-1=0 \quad \text{and} \quad
\beta^{p}\alpha^{jp}-1=0\; .
$$
Therefore, $ \beta^{p}\alpha^{ip}=\beta^{p}\alpha^{jp} $, and
$\alpha^{ip}=\alpha^{jp}$, hence,
$\alpha^{(i-j)p}=1$. For any $i\neq j , i,j=0,\ldots,n-1$, this
can be true only if $\gcd(p,n)>1$.
\end{IEEEproof}

\subsection{Known Bounds On the Minimum Distance}
Let us shortly recall well-known bounds on the minimum distance of cyclic codes.
\begin{theorem}[Hartmann--Tzeng (HT) Bound,~\cite{Hartmann_GeneralizationsofBCHbound_1972}] \label{def_BCHCode}
Let \CYC{\Fq}{n}{k}{d} be a $q$-ary cyclic code of length $n$, dimension $k$, distance $d$ and with defining set $D_{\mathcal{C}}$. Let
\begin{equation*}
\begin{split}
 & \{ b + i_1 \HTa +i_2 \HTb, \; \forall i_1 =0,\dots,\muHT - 2, i_2 = 0,\dots,\nuHT \} \\
 & \subseteq D_{\mathcal{C}},
\end{split}
\end{equation*}
where $\gcd(n,\HTa) = 1$ and $\gcd(n,\HTb) = 1$. Then $d \geq \dHT \defeq \muHT+\nuHT$.
\end{theorem}
Note that for $\nuHT = 0$ the HT bound becomes the BCH bound~\cite{Hocquenghem_1959, Bose_RayChaudhuri_1960} and it is denoted by \dBCH.
A further generalization was proposed by Roos~\cite{Roos_GeneralBCHBound_1982, Roos_BoundforCyclicCodes_1983}.

\subsection{BCH Bound with Rational Function}
Let $c(x) = \sum_{i=0}^{n-1} c_i x^i$ denote the polynomial representation of a codeword $\cw$ of a
cyclic code \CYC{\Fq}{n}{k}{d \geq \muHT}. We consider the BCH bound in the following and assume that $\nuHT = 0$ and $\HTa=1$
 and therefore $c(\alpha^{i}) = 0, \; \forall i =b, \dots, b+ \muHT -2 $, such that $\muHT$ is maximal. Let
the formal power series $a(b, \alpha^i x)$
\begin{equation} \label{eq_simple_frac}
a(b,\alpha^i x) \defeq \frac{\alpha^{ib}}{1-\alpha^i x} = \alpha^{ib} \sum_{j=0}^{\infty} (\alpha^i x)^j
\end{equation}
be given.
For any $c(x) \in \CYC{\Fq}{n}{k}{d}$ we can rewrite the BCH bound as follows:
\begin{align} \label{eq_simple_power}
\sum_{j=0}^{\infty} c(\alpha^{j+b}) x^{j} & = \sum_{i=0}^{n-1} c_i \alpha^{ib} + \sum_{i=0}^{n-1} c_i \alpha^{i} \alpha^{ib} x + \dots \nonumber  \\
 & \equiv 0 \bmod x^{\muHT-1},
\end{align}
and with~\refeq{eq_simple_frac} we can rewrite~\refeq{eq_simple_power} as:
\begin{align} \label{eq_simple_bch}
\sum_{i=0}^{n-1} c_i \frac{\alpha^{ib}}{1-\alpha^i x} &  = \sum_{i=0}^{n-1} c_i \cdot a(b,\alpha^ix) \nonumber \\
& \equiv 0 \bmod x^{\muHT-1}.
\end{align}
Let $\mathcal W$ be the set of nonzero positions of a
codeword and let $|\mathcal W| = d $.
With $\gcd(1-\alpha^ix, 1-\alpha^jx) = 1, \; \forall i \neq j$, we can write~\refeq{eq_simple_bch} as follows:
\begin{equation} \label{eq_BCH_Rational}
\frac{\sum \limits_{i \in \mathcal W} \Big(c_i \cdot \alpha^{ib} \cdot \prod_{\substack{j \in \mathcal W\\ j \neq i}} (1-\alpha^j x) \Big)}{\prod_{i \in \mathcal W} (1-\alpha^i x) } \equiv 0  \bmod x^{\muHT-1},
\end{equation}
where the degree of the numerator is less than or equal to $d-1$ and
has to be greater than or equal to $\muHT-1$ to obtain zero on the RHS of~\refeq{eq_BCH_Rational}. Then, the minimum distance $d$ of a cyclic code $\mathcal{C}$ is $d \geq \muHT$.

\section{Roots of Cyclic Codes Represented by Rational Functions} \label{sec_principle}
Our idea for bounding the distance of $q$-ary cyclic codes
originates from the definition and properties of classical
Goppa codes~\cite{Goppa_RationReprCodes_1971 ,
Goppa_NewClassCodes_1970 } and generalized Goppa
codes~\cite{Bezzateev_OneGenGoppaCodes_1997
,Shekhunova_LGCodes_1981}. We do not present
the theory of Goppa codes here, since we use only the properties of rational
functions introduced in Section~\ref{sec_preliminaries}.

Let $b$ be an integer and let $\alpha$ be an $n$th root of unity.
Let $h(x), f(x) \in \Fxq$ with degree $v$ and $u$ and with $\deg \gcd (h(x),f(x)) = 0$ be given.
The power series $ a(b, \alpha^{i} x)$ is defined such that:
\begin{align} \label{eq_powerseries_def}
 a(b, \alpha^{i} x) & \defeq \frac{\alpha^{ib} h(\alpha^i x)}{f(\alpha^i x)} = \sum_{j=0}^{\infty} a_j \alpha^{ib}(\alpha^{i}x)^j \nonumber \\
& = a_0 \alpha^{ib} + a_1 \alpha^{ib}\alpha^{i} x + a_2
\alpha^{ib}(\alpha^{i} x)^2 + \dots \; .
\end{align}

Similar to the case of the BCH bound, we associate a $q$-ary cyclic code $\mathcal{C}$ with a power series $a(b, \alpha^{i} x)$ as follows.

\begin{definition}[Connection between Power Series and
Code] \label{def_PowerSeriesCode}
Let a power series  $a(b, \alpha^{i} x)$ (or respectively two polynomials $h(x),f(x)$ and an integer $b$) with
 $\deg \gcd(h(x),f(x)) = 0$ and a $q$-ary cyclic code $\CYC{\Fq}{n}{k}{d}$ be given. Furthermore, let $\gcd\big(n, p(h(x)/f(x))\big) =1$. Let $\alpha$ denote an $n$th root of unity. Then, there exist a $\mu \geq 0$, such that for all $c(x) \in \mathcal{C}$:
\begin{equation} \label{eq_powerseriesandcodeword}
\begin{split}
\sum_{j=0}^{\infty} a_j c(\alpha^{j+b}) x^j & \equiv 0 \bmod x^{\mu-1}
\end{split}
\end{equation}
holds.
\end{definition}
Before we prove the main theorem on the minimum distance of a cyclic code $\mathcal C$, let us describe Definition~\ref{def_PowerSeriesCode}.
We search the longest ``sequence``
\begin{equation*}
a_0 c(\alpha^b) , a_1 c(\alpha^{b+1}) , \dots , a_{\mu-2} c(\alpha^{b+\mu-2}),
\end{equation*}
that is a zero-sequence, i.e., the product of the coefficient $a_j$ and the evaluated codeword $c(\alpha^{b+j})$ gives zero for all $j=0,\dots,\mu-2$.
We require a root $\alpha^{j}$ of the code $\mathcal C$, if the coefficient $a_{j-b}$ of the power series $a(b, \alpha^{i} x)$ is nonzero.

Equation~\refeq{eq_powerseriesandcodeword} can be rewritten in terms of the polynomials 
$h(x)$ and $f(x)$ as follows:
\begin{align} \label{eq_frac_further}
\sum_{j=0}^{\infty} a_j c(\alpha^{j+b}) x^j & = \sum_{j=0}^{\infty} \sum_{i=0}^{n-1} a_j c_i \alpha^{i(j+b)} x^j \nonumber \\
& = \sum_{i=0}^{n-1} c_i \Big( \sum_{j=0}^{\infty} a_j \alpha^{i(j+b)} x^j   \Big) \nonumber \\
& = \sum_{i=0}^{n-1} c_i \frac{\alpha^{ib} h(\alpha^i x)}{f(\alpha^i x)} \nonumber \\
& \equiv 0 \bmod x^{\mu-1}.
\end{align}
Let $\mathcal W$ be the set of nonzero positions of a
codeword and let $|\mathcal W| = d $.
With $\deg \gcd(f(\alpha^i x), f(\alpha^j x)) = 0, \; \forall i \neq j$ (that follows from $\gcd\big(n, p(h(x)/f(x))\big)=1$ according to Lemma~\ref{lem_coprime}), we can write~\refeq{eq_frac_further} as
\begin{align} \label{eq_frac_commondenominator}
& \frac{ \sum \limits_{i \in \mathcal W} \Big(c_i \cdot \alpha^{ib} \cdot h(\alpha^{i}x) \cdot \prod_{\substack{j \in \mathcal W\\ j \neq i}} f(\alpha^j x) \Big)} {\prod_{i \in \mathcal W} f(\alpha^{i}x )} \nonumber \\
& \equiv 0 \bmod x^{\mu-1},
\end{align}
where the degree of the denominator is $u d$ and the numerator has
degree smaller than or equal to $(d-1)u+v$. This leads to
the following theorem on the minimum distance of $\mathcal{C}$.

\begin{theorem}[Minimum Distance] \label{lem_minimumdistance}
Let a $q$-ary cyclic code \CYC{\Fq}{n}{k}{d} be given and let
$\alpha$ denote an $n$th root of unity. Let two co-prime
polynomials $h(x)$ and $f(x)$ in $\Fxq$ with degrees $v$
and $u$, respectively and the integers $b$ and $\mu$ be given, such
that~\refeq{eq_frac_commondenominator} holds. Let $\gcd\big(n, p(h(x)/f(x))\big)=1$.

Then, the minimum distance $d$ of \CYC{\Fq}{n}{k}{d} satisfies the following
inequality:
\begin{equation}\label{inequality_for_d}
d \geq d_f \defeq \left \lceil \frac{\mu-1-v}{u}+1 \right \rceil.
\end{equation}
\end{theorem}
\begin{IEEEproof}
For a codeword $\cw$ of weight $d$, the degree of the numerator
in~\refeq{eq_frac_commondenominator} is less than or equal to
$(d-1)u+v$ and has to be greater than or equal to $\mu-1$.
\end{IEEEproof}

\begin{example}[Binary Cyclic Code] \label{ex_fraction}
Consider the binary cyclic code \CYC{\F{2}}{17}{9}{5} with defining set $D_{\mathcal{C}} = M_1 =\{ 1,2,4,8,16,15,13,9\} \equiv \{1,2,4,8,-1, -2, -4 , -8 \} \mod 17 $.
Let $b=-4, h(x)=x+1$ and $f(x)=x^2+x+1 \in \Fx{2}$ be given.
Then, $a(-4, \alpha^ix)$ has according to Definition~\ref{def_periodrational} period of three and we have
$(a_0 \ a_1 \ a_{2}) = (1 \ 0 \ 1)$.

The following table illustrates how we match the roots of the generator polynomial to the zeros
of the power series expansion $a(-4, \alpha^ix)$.
In the first row, the defining set is shown, i.e., $c(\alpha^j)=0$
for all $j \in \defset$.
The $\square$ marks elements that are not necessarily roots of the code.
In the second row of the table, the power series expansion $\mathbf a = (a_0 \ a_1 \ a_2 \ a_0 \ a_1 \ \dots)$ is shown
for the considered interval.
\medskip
\begin{center}
 \begin{footnotesize}
  \begin{tabular}{l||*{8}{c|}c}
   $\defset $ &  -$4$ & $\square$ &-$2$ & -$1$& $\square$ & $1$ & $2$& $\square$ & $4$ \\
   \hline
   $\mathbf a $ & $1$ & $0$ & $1$ & $1$ & $0$ & $1$ & $1$ & $0$ & $1$\\
   \end{tabular}
 \end{footnotesize}
\end{center}
We have $a_{j} \cdot c(\alpha^{j-4})=0, \, \forall j = 0,\dots, 8$, for all
$c(x) \in \CYC{\F{2}}{17}{9}{5}$.
We obtain a zero-sequence of length $\mu-1 = 9$ and therefore with Theorem~\ref{lem_minimumdistance}, $d_f = 5$. This is the actual distance $d$ of this code.

In next section, we see that $\CYC{\F{2}}{17}{9}{5}$ belongs to the class of reversible codes and we
can associate this rational function to the whole class.
\end{example} 
Let us illustrate the case where $\deg h(\alpha^{i} x) > 0$.
For $h(\alpha^{i} x) =  h_0 + h_1 \alpha^{i} x + \dots + h_v(\alpha^{i} x)^v$ we decompose
the power series expansion of~\refeq{eq_powerseries_def} into:
\begin{equation} \label{eq_powerseries_linearcombination}
a(b, \alpha^{i} x) = \alpha^{ib} \left( \frac{h_0}{f(\alpha^{i}x)} + \dots +\frac{h_{v}(\alpha^{i}x)^v}{f(\alpha^{i}x)}  \right).
\end{equation}
Our classification of $q$-ary cyclic codes based on Theorem~\ref{lem_minimumdistance} works as follows. In the first step, we
consider the power series expansion $1/f(x) = (\bar{a}_0 + \bar{a}_1 x + \dots + \bar{a}_{p-1}x^{p-1})/(-x^p+1) $ with period $p= p(1/f(\alpha^{i}x))$.
From~\refeq{eq_powerseries_linearcombination} we can interpret $a(b, \alpha^{i} x)$ as a linear combination of $v+1$ shifted series expansion $1/f(\alpha^{i}x)$:
\begin{align*}
   & h_0 (\bar{a}_0 \ \bar{a}_1 \ \dots \ \bar{a}_{p-1})  \\
  + \; &  h_1 (\bar{a}_{p-1} \  \bar{a}_0 \ \dots \ \bar{a}_{p-2}) \\
  + \; & \quad \quad \quad \quad \quad \vdots \\
  + \; & h_v (\bar{a}_{p-v} \ \bar{a}_{p-v+1} \ \dots \ \bar{a}_{p-1-v}) \\
  = \; &  (a_0 \ a_1 \ \dots \ a_{p-1}).
\end{align*}
Then, we can select $b$ such that the characteristic sequence of $a_{0} c(\alpha^b),a_{1} c(\alpha^{b+1}),\dots, a_{\mu-2} c(\alpha^{b+\mu-2})$ becomes zero for the maximal $\mu$ of a given code $\CYC{\Fq}{n}{k}{d}$.

\section{On the Distance of Some Classes\\ of q-ary Cyclic Codes} \label{sec_classes}
\subsection{Structure of Classification and Cardinality}
Before we describe our classification let us extend Definition~\ref{def_PowerSeriesCode}. We introduce an equivalent parameter to $m_1$ and $m_2$ of the HT bound which is denoted by $\fracjump$.
We search for a given power series $a(b, \alpha^{i} x)$ and a cyclic code $\mathcal{C}$ the "longest" sequence:
\begin{equation*}
a_0 c(\alpha^b) , a_1 c(\alpha^{b+\fracjump}) , \dots , a_{\mu-2} c(\alpha^{b+(\mu-2)\fracjump}),
\end{equation*}
that is a zero-sequence of length $\mu-1$.

We classify $q$-ary cyclic codes by subsets of their defining
set $\defset$ and their length $n$. We specify our new lower bound (Theorem~\ref{lem_minimumdistance}) on the
minimum distance for some classes of codes. Additionally, we compare it to the BCH~\cite{Hocquenghem_1959, Bose_RayChaudhuri_1960} and the HT\cite{Hartmann_GeneralizationsofBCHbound_1972} bound, which we denote by $\dBCH$ and $\dHT$.

We use the following power series expansions $1/f(x)$ over $\Fq$ with period $p$,
where $\mathbf a =  (a_0 \ a_1 \ \dots \ a_{p-1})$ denotes the coefficients.
\begin{itemize}
\item $1/(x^2+x+1)$ over $\F{q}$ \\ with $\mathbf a = (1 \ $-$1 \ 0 )$ and $p=3$,
\item $1/(x^3+x^2+x+1)$ over $\F{q}$ \\ with $\mathbf a = (1 \ $-$1 \ 0 \ 0)$ and $p=4$,
\item $1/(x^3+x+1)$ over $\F{2}$ \\ with $\mathbf a = (1 \ 1 \ 1 \ 0 \ 1 \ 0 \ 0)$ and $p=7$,
\item $1/(x^4+x+1)$ over $\F{2}$ \\ with $\mathbf a = (1 \ 1 \ 1 \ 1 \ 0 \ 1 \ 0 \ 1 \ 1 \ 0 \ 0 \ 1 \ 0 \ 0 \ 0)$ and $p=15$.
\end{itemize}
We match a power series expansion $a(b,\alpha^ix)$
to the roots of the generator polynomial, such that
$a_j \cdot g(\alpha^{b+j\fracjump}) = a_j \cdot c(\alpha^{b+j\fracjump}) = 0, \, \forall j=0,\dots,\mu-2$.

Throughout this section, we assume due to Lemma~\ref{lem_coprime}
that $\gcd(n,p)=1$ and we use Theorem~\ref{lem_minimumdistance}
to state the lower bound $d_f$ on the
distance of the codes.

In Table~\ref{tab_denominators}, all cyclic shifts of the power series expansions of $1/(x^2+x+1)$ and $1/(x^3+x^2+x+1)$ are
shown and the corresponding numerator $h(x)$ is given.
\begin{table}[htb]
    \caption{Power series $(a_0 \ \dots \ a_{p-1})$ for the rational functions $1/(x^2+x+1)$ and  $1/(x^3+x^2+x+1)$ and their corresponding cyclic shift.}
    \label{tab_denominators}
    \centering
      \setlength{\extrarowheight}{\defrowheight}
      \begin{tabular}{c|c|c}
     $(a_0 \ \dots \ a_{p-1})$&$ f(x)$ & $ h(x)$ \\
     \hline
     \hline
     $(1 \ $-$1 \ 0)$ & $1+x + x^2$ & $1$\\
     $($-$1 \ 0 \ 1)$ & $1+x + x^2$ & $ -1- x$\\
     $(0\ 1\ $-$1)$ & $1+x + x^2$ & $x$\\
     \hline
     $(1 \ $-$1 \ 0 \ 0)$ & $ 1+x + x^2 + x^3 $ & 1\\
     $(0 \ 1 \ $-$1 \ 0)$ & $ 1+x + x^2 + x^3 $ & $x$\\
     $(0 \ 0 \ 1 \ $-$1)$ & $ 1+x + x^2 + x^3 $ & $x^2$\\
     $($-$1 \ 0 \ 0 \  1)$ & $ 1+x + x^2 + x^3 $ & $ -1 -x -x^2 $\\
    \end{tabular}
\end{table}
First, we apply our approach to the wide class of reversible codes.
Afterwards, we show how our principle can equivalently be used for non-reversible
codes.

\subsection{Reversible Codes}
In this subsection, we show how our approach can be applied for a large class of cyclic codes --- the class of \emph{reversible codes} \cite{Massey_ReversibleCodes_1964, MacWilliamsSloane_TheTheoryOfErrorCorrecting_1988}. A code $\mathcal C$ is reversible if for any codeword $\c = \cw \in \mathcal C$ also
$\c = (c_{n-1} \ c_{n-2}\ \dots \ c_0) \in \mathcal C$. A cyclic code is reversible if and only if the reciprocal of every zero of the generator polynomial $g(x)$ is also a zero of $g(x)$, i.e.,
\begin{equation}
\defset = \lbrace i_1,i_2,\dots,i_{\ell},-i_1,-i_2,\dots,-i_{\ell}\rbrace.
\end{equation}
A special class of reversible codes, which we call \emph{symmetric reversible codes} is given based on the following lemma.
\begin{lemma}[Symmetric Reversible Codes]\label{lem:symmreversible}
Let $n$ be the length of a $q$-ary cyclic code.
Any union of cyclotomic cosets is a defining set of a reversible code if and only if $n \mid (q^m+1)$, for some $m \in \mathbb N$.
\end{lemma}
\begin{IEEEproof}
Any union of cyclotomic cosets defines a reversible code if and only if any coset is reversible, i.e., if for all $r$
and some integer $m$:
\begin{equation*}
M_r = \lbrace r, r \cdot q, \dots, r\cdot q^{m-1},-r,-r\cdot q, \dots,-r\cdot q^{m-1}  \rbrace.
\end{equation*}
Therefore for all $r$, the following has to hold:
\begin{equation*}
r \cdot q^m \equiv -q \mod n \quad \Longleftrightarrow \quad r \cdot (q^m+1) \equiv 0 \mod n.
\end{equation*}
Since $r=1$ always defines a cyclotomic coset, $(q^m+1) \equiv 0 \mod n$ has to hold.
This is fulfilled if and only if $n \mid (q^m+1)$ and in this case also $r \cdot (q^m+1) \equiv 0 \mod n$ holds for any $r$.
\end{IEEEproof}

Moreover, the following lemma provides the cardinality of all cyclotomic cosets if $n \mid (q^m+1)$.
\begin{lemma}[Cardinality of Symmetric Reversible Codes]
Let $m$ be the smallest integer such that $n$ divides $(q^m+1)$,
then the cardinality of the cyclotomic coset $M_r$ is $|M_r|=2m$ if $\gcd(n,r)=1$.
\end{lemma}
\begin{IEEEproof}
Since $n \mid (q^{m}+1)$, it follows also that $n \mid (q^{m}+1)(q^{m}-1) = (q^{2m}-1)$.
Since $m$ is the smallest integer such that $n$ divides $(q^m+1)$, also $s \defeq 2m$ is
the smallest integer such that $n \mid (q^{s}-1)$.
With Lemma~\ref{lem:cardinality_general}, we obtain $|M_r|=s$ if
$\gcd(n,r)=1$. 
Therefore, $|M_r|=s= 2m$.
\end{IEEEproof}
In order to illustrate our bound, we first restrict ourselves to binary codes.
To give a new bound on the minimum distance,
we first use the rational function $a(x) = h(x)/f(x)$ with $f(x) = x^2+x+1$, where $p(a(x))=3$.
\begin{table}[htb]
    \caption{Bounds on the distance of $q$-ary cyclic codes of length $n\mid (q^{s}-1)$ and $\gcd(n,3) = 1$, using $f(x)=x^2+x+1$}
    \label{tab_paramall}
    \centering
	\setlength{\extrarowheight}{\defrowheight}
	\begin{tabular}{c|c c c}
     Binary&&&\\ [-1.5ex]
     Symmetric&$ \{ 1\} \subseteq \defset $ &$\{ 1,5 \} \subseteq \defset$ & $\{ 1,5,7 \} \subseteq \defset$ \\ [-1ex]
     Reversible&&&\\
     &$k \geq n-\ell$ & $k \geq n-2\ell$ & $k \geq n-3\ell$\\
     \hline
     Binary &$ \{$-$1, 1\}\subseteq \defset $ &$\{$-$5,$-$1, 1,5 \}$ & $\{$-$7,$-$5,$-$1,1,$\\[-1ex]
     Reversible &&$\subseteq \defset$& $5,7 \}\subseteq \defset$\\
     &$k \geq n-2\ell$&$k \geq n-4\ell$&$k \geq n-6\ell$\\
     \hline
     General &$\{ $-$4,$-$2,$-$1,1,$&$\{ $-$5,$-$4,$-$2,$-$1,1,$&$\{$-$10,$-$7,$-$5,$-$4,$-$2,$\\[-1ex]
     $q$-ary&$2,4\} \subseteq \defset$& $2,4,5 \} \subseteq \defset$&-$1,1,2,4,5,$\\[-1ex]
     &&&$7,10\} \subseteq \defset$\\
     \hline
     BCH&$ \dBCH=4$&$ \dBCH=5$&$\dBCH=8$\\
     &$b=-4$&$b=-5$&$b=-10$\\
     &$\HTa=3$&$\HTa=3$&$\HTa=3$\\
     \hline
     HT&$\dHT=5$&$\dHT=6$&$\dHT=9$\\
     &$b=-4$&$b=-5$&$b=-10$\\
     &$\HTa=3$&$\HTa=3$&$\HTa=3$\\[-1ex]
     &$\HTb = 2$&$\HTb = 1$&$\HTb = 2$\\
     &$\muHT = 4$, $\nuHT=1$&$\muHT = 5$, $\nuHT=1$&$\muHT = 8$, $\nuHT=1$\\
     \hline
     Fractions&$d_f=5$&$d_f=7$&$d_f=11$\\
     &$b=-4$&$b=-6$&$b=-10$\\[-1ex]
     &$\fracjump=1$&$\fracjump=1$&$\fracjump=1$\\[-1ex]
     &$\mu=10$&$\mu=14$&$\mu=22$\\[-1ex]
     &$\mathbf a = (-1 \ 0\ 1)$&$\mathbf a = (0 \ 1 \ -1)$&$\mathbf a = (-1 \ 0\ 1)$
    \end{tabular}
\end{table}
For a binary symmetric reversible code $\mathcal C$, we showed that each cyclotomic coset
is symmetric. Therefore, if $\{ 1\} \subseteq \defset $, we
know that $\{-4,-2,-1,1,2,4\}$ is in the defining set.
Let us use the (cyclically shifted) power series expansion $\mathbf a = (-1 \ 0 \ 1 \ \dots)$.
According to Table~\ref{tab_denominators}, we have $h(x) = -1 -x$.
We match the roots of $\mathcal C$ for $b=-4$ and $\fracjump=1$, to a zero-sequence of length $\mu-1=9$.
Therefore our bound provides $d \geq d_f =5$.

Let the defining set $\defset$ of the binary symmetric reversible code $\mathcal C$
additionally include $5$.
Then we obtain for $b=-6$ and $\fracjump = 1$ a sequence of length $\mu-1=13$, which results in $d_f = 7$.

In the same way, if $\{ 1,5,7 \} \subseteq \defset$, we obtain $\mu-1=21$ with
$b = -10$ and $\fracjump=1$ and thus, $d_f = 11$.
These parameters are shown in Table~\ref{tab_paramall} and compared with the BCH and HT bounds.

As mentioned before, reversible codes are defined such that the reciprocal of each root
of the generator polynomial is also a root. Therefore, a
defining set where $ r \subseteq \defset$, and also $-r \subseteq \defset$ defines a
reversible code if $\gcd(r,n)=1$ and $\gcd(-r,n)=1$. The conditions are
necessary to guarantee that both cyclotomic cosets have the same cardinality
(compare Lemma~\ref{lem:cardinality_general}) and hence each reciprocal root is
also in the defining set.
The second row of Table~\ref{tab_paramall} shows which subsets have to be in the defining
set in order to obtain the same parameters as for binary \emph{symmetric} reversible codes.
Note that
$s$ is the smallest integer such that the length $n$ divides $q^{s}-1$.

This principle can easily be generalized to $q$-ary codes.
The third row of Table~\ref{tab_paramall} gives these results in general.
Note that in Table~\ref{tab_paramall}, $\gcd(n,p = 3)=1$ has
to hold because of Lemma \ref{lem_coprime}.
\begin{example}[Binary Symmetric Reversible Code]
The binary cyclic code \CYC{\F{2}}{17}{9}{5} from Example~\ref{ex_fraction} is symmetric reversible
since Lemma~\ref{lem:symmreversible} is fulfilled.
If $\{1\} \subseteq \defset$, then $\defset = \{ 1,2,4,8,16,15,13,9\} \equiv \{1,2,4,8,-1, -2, -4 , -8 \} \mod 17 $ and we obtain $d_f = 5$.

For this class of binary cyclic codes, the bound $d \geq 5$ on the minimum distance can be also obtained by another way (as pointed out by a reviewer). With $b=-4$ and $\HTa=3$ we know from the BCH bound that the minimum distance is at least four. A binary cyclic code of even weight codewords has the zero in the defining set and we would obtain five consecutive zeros (resulting in a minimum distance of at least six). This implies that a codeword of weight four can not exists and therefore a binary cyclic code $\mathcal C $, where $\{-4,-2,-1,1,2,4\}
\subseteq \defset $, has at least minimum distance five.
\end{example}
In Table~\ref{tab_QaryPeriod4}, we list some classes of cyclic codes where the denominator $f(x)$ of
the rational function $\alpha^{ib}h(\alpha^ix)/f(\alpha^i x)$ has degree three and the period
is $p(1/(x^3+x^2+x+1)) = 4$.
\begin{table}[htb]
    \caption{Bounds on the distance of $q$-ary cyclic codes of length $n|(q^{s}-1)$ and $\gcd(n,4) = 1$, using $f(x)=x^3+x^2+x+1$.}
    \label{tab_QaryPeriod4}
    \centering
	\setlength{\extrarowheight}{\defrowheight}
	\begin{tabular}{c|c c c}
     Binary&&&\\[-1.5ex]
     Symmetric&$ \{3,5\} \subseteq \defset $ &$\{ 3,5,11 \}\subseteq \defset$ & $\{ 3,5,11,$ \\[-1.5ex]
     Reversible&&& $13 \} \subseteq \defset$\\
     &$k \geq n-2\ell$&$k \geq n-3\ell$&$k \geq n-4\ell$\\
     \hline
     Binary &$ \{$-$5,$-$3,3,$ &$\{$-$11,$-$5,$-$3,$ & $\{$-$13,$-$11,$-$5,$-$3,3,$\\[-1ex]
     Reversible & $ 5\} \subseteq \defset$ & $3,5,11 \} \subseteq \defset$& $ 5,11,13 \}\subseteq \defset$\\
     &$k \geq n-4\ell$&$k \geq n-6\ell$&$k \geq n-8\ell$\\
     \hline
     General &$ \{$-$6,$-$5,$-$3, $ &$\{$-$11,$-$6, $-$5,$ & $\{$-$13,$-$11,$-$6,$-$5,$\\[-1ex]
     $q$-ary & $ 3,5,6\}$ & $$-$3,3,5,6,$& $ $-$3,3,5,6, $ \\[-1ex]
     & $\subseteq \defset$ & $ 11 \} \subseteq \defset $ & $11,13 \}\subseteq \defset$\\
     \hline
     BCH & $\dBCH=3$ & $\dBCH=3$ & $\dBCH=4$\\
     & $b=-6$ & $b=-6$ & $b=-13$\\[-1ex]
     & $\HTa=1$ & $\HTa=1$ & $\HTa=1$\\
     \hline
     HT & $\dHT=\dBCH$ & $\dHT=5$ & $\dHT=6$\\
     & $b=-6$ & $ b=-11 $ & $b=-13$\\[-1ex]
     & $\HTa=1$ & $\HTa=8$ & $\HTa=8$\\[-1ex]
     & $\HTb = 0$ & $\HTb = 6$ & $\HTb = 2$\\
     & $\muHT = 3$ & $\muHT = 4$, $\nuHT=1$ & $\muHT = 5$, $\nuHT=1$\\
     \hline
     Fractions & $d_f=4$ & $d_f=5$ & $d_f=7$\\
     & $b=-9$ & $b=-11$ & $b=-17$\\[-1ex]
     & $\fracjump=2$ & $\fracjump=2 $ & $\fracjump = 2$\\[-1ex]
     &$\mu=11$&$\mu=13$&$\mu=19$\\
     & $\mathbf a = (0 \ 0\ 1 \ $-$1)$ & $\mathbf a = (0 \ 0\ 1 \ $-$1)$ & $\mathbf a = (0 \ 0\ 1 \ $-$1)$
    \end{tabular}
\end{table}
The power series expansion is $1/(x^3+x^2+x+1) = (1-x)/(-x^4+1)$.
Let us consider the second class, where in the case of a binary symmetric reversible code the set $\{ 3,5,11 \}$ must be 
in the defining set of the code. The HT bound gives the same lower bound on the minimum distance as our approach $\dHT = 5$.

\begin{example}[Binary Cyclic Code]
The binary cyclic code \CYC{\F{2}}{45}{31}{4} with $ \defset = \{ -5,-3,3,5 \} = \{3,5,6,10,12,20,21,24,25,33,35,39,40,42\}$ is in the class of codes in the first
column of Table~\ref{tab_QaryPeriod4}. We obtain $d_f =4$, which is the actual distance
of the code.

Note that $3 \mid 45$ and therefore we can not use Table~\ref{tab_paramall}.
\end{example}

\subsection{Non-Reversible Codes}
In this subsection, we show that our principle equivalently can be used for non-reversible codes.
We use one $f(x)$ of degree three and one $f(x)$ of degree four.
We give some classes of binary cyclic codes in this subsection to show
the principle.
The power series expansion of the polynomial $f(x)=x^3+x+1$ over $\Fx{2}$ has
period $p=7$.
To obtain a bound on the minimum distance, we consider
the case of extended binary cyclic codes, where the $0$ is in
the defining set $\defset$.
Assume that $\{ -3,0,1, 7\} \subseteq \defset$.
The sequence of zeros of the binary code
can be matched to the rational function for $b=-4$ and $\fracjump=1$.
The corresponding distance is then $d_f = 5$.
This and some other combinations of subsets of $\defset$ are shown in Table~\ref{tab_paramDegThree}.
\begin{table}[htb]
    \caption{Bounds on the distance of binary cyclic codes of length $n\mid (2^{s}-1)$
    and $\gcd(n,7) = 1$, using $f(x)=x^3+x+1$}
    \label{tab_paramDegThree}
    \centering
    \setlength{\extrarowheight}{\defrowheight}\begin{tabular}{c|ccc}
    Binary&$\{-3, 0,1,7\}$&$\{-3, 0,1,7,9\}$ &$\{-3, 0,1,7,9,11\}$\\[-1ex]
    Codes&$\subseteq \defset$&$\subseteq \defset$&$\subseteq \defset$\\
    &$k \geq n-4\ell$&$k \geq n-5\ell$&$k \geq n-6\ell$\\
    \hline
     BCH&$\dBCH=4$&$\dBCH=4$&$\dBCH=4$\\
     &$b=-3$&$b=-3$&$b=-3$\\[-1ex]
     &$c_1=5$&$c_1=5$&$c_1=5$\\
     \hline
     HT&$\dHT=4$&$\dHT=4$&$\dHT=4$\\
     &$b=-3$&$b=-3$&$b=-3$\\[-1ex]
     &$\HTa=5$&$\HTa=5$&$\HTa=5$\\[-1ex]
     &$\HTb = 0$&$\HTb = 0$&$\HTb = 0$\\[-1ex]
     &$\muHT = 4$, $\nuHT=0$&$\muHT = 4$, $\nuHT=0$&$\muHT = 4$, $\nuHT=0$\\
     \hline
     Fractions&$d_f=5$&$d_f=6$&$d_f=7$\\
     &$b=-4$&$b=-4$&$b=-4$\\[-1ex]
     &$\fracjump=1$&$\fracjump=1$&$\fracjump=1$\\[-1ex]
     &$\mu=14$&$\mu=16$&$\mu=19$\\[-1ex]
     &$\mathbf a = $&$\mathbf a = $&$\mathbf a = $\\[-1.5ex]
     &$(1 \ 0 \ 0 \ 1 \ 1 \ 1 \ 0 )$&$(1 \ 0 \ 0 \ 1 \ 1 \ 1 \ 0 )$&$(1 \ 0 \ 0 \ 1 \ 1 \ 1 \ 0 )$\\
    \end{tabular}
\end{table}
Another class of binary cyclic codes can be identified using the polynomial $f(x)=x^4+x+1$ with $p(1/f(x)) = 15$.
We use the shifted power series expansion such that $\mathbf a = (1 \ 0\ 0\ 1\ 0\ 0\ 0\ 1\ 1\ 1\ 1\ 0\ 1\ 0\ 1)$.

As required by Lemma~\ref{lem_coprime}, we only consider lengths $n$, such that $\gcd(n,p = 15)=1$.
We can match a concatenation of $\mathbf a$
to the roots of the generator polynomial for $b=-6$ and $\fracjump=1$ if $\{ 1,3,9,-3\}\subseteq \defset$.
Our bound on the distance yields
$d_f = 6$, since $\deg f(x) = 4$, whereas the BCH and the HT bound give $\dBCH = \dHT = 5$.

Table~\ref{tab:boundscodesq2} and~\ref{tab:boundscodesq3} in the appendix
show our bound for binary and ternary cyclic codes. We used the power series expansions of $1/(x^2+x+1)$ and
$1/(x^3+x^2+x+1)$ to obtain a good refinement of our new bound on the minimum distance.
We list the number of codes, for which the BCH bound is not tight ($\#\dBCH < d$), the number
of cases, where our bound is better than the BCH bound ($\#d_{f} > \dBCH$) and count the cases, where our bound is not
tight ($\#d_f < d$).
All lengths $n$, for which any union of cyclotomic cosets is a
symmetric reversible code, are marked by a star $*$.

\section{Generalizing Boston's Bounds}\label{sec_boston}
In~\cite{Boston_CyclicCodesAlgebraicGeometry_2001}, Boston gave ten bounds, denoted by $\dBOSTON$, on
the minimum distance of $q$-ary cyclic codes, which he proved using algebraic geometry.
These bounds are each for a specific subset of the defining set and do not
consider whole classes of codes.
In this section, we show how our approach generalizes some of these bounds.

Six of Boston's ten bounds are given as follows.
\begin{theorem}[Boston Bounds,~\cite{Boston_CyclicCodesAlgebraicGeometry_2001}]
The following bounds on the minimum distance of a $q$-ary cyclic code $\mathcal C$ hold:
\begin{itemize}
 \item[1)] If $3 \nmid n$ and $\{ 0,1,3,4\} \subseteq \defset$, then $\dBOSTON = 4$,
 \item[2)] If $\{ 0,1,3,5\} \subseteq \defset$, then $\dBOSTON = 4$,
 \item[5)] If $3 \nmid n$ and $\{ 0,1,3,4, 6\} \subseteq \defset$, then $\dBOSTON = 5$,
 \item[6)] If $4 \nmid n$ and $\{ 0,1,2,4,5, 6,8\} \subseteq \defset$, then $\dBOSTON = 6$,
 \item[7)] If $3 \nmid n$ and $\{ 0,1,3,4, 6, 7\} \subseteq \defset$, then $\dBOSTON = 6$,
 \item[10)] If $3 \nmid n$ and $\{ 0,1,3,4, 6, 7,9\} \subseteq \defset$, then $\dBOSTON = 7$.
\end{itemize}
\end{theorem}
We use again two power series expansions $1/f(x)$.
The first power series expansion is $1/(x^2+x+1)$ of
period $p=3$ with $(a_0 \ a_1 \ a_2)=(1 \ $-$1 \ 0)$.
The second considered power series expansion $1/(x^2+1)$ has period $p=4$ with
$(a_0 \ a_1 \ a_2 \ a_3)=(1 \ 0 \ $-$1 \ 0)$. Note that the latter is
actually a special case of the BCH bound.
Table~\ref{tab_bostonbounds} shows the six Boston bounds. Boston's bounds 1,2,5,6 and 7 are special cases of our bounds.
However, for Boston's bound 10, our approach gives a worse bound.
\begin{table}[htb!]
    \caption{Boston's Bounds}
    \label{tab_bostonbounds}
    \centering
	\setlength{\extrarowheight}{\defrowheight}
	\begin{tabular}{c|c|c|c|c|c}
     No &$\mathcal I = $ &$f(x)$&$\mathbf{a}$ & $d_f$& Conditions\\
     \hline
     1	&$[$-$1,5]$&$x^2+x+1$	&$(0\ 1\ $-$1 \dots)$	&$4$	&$\gcd(n,3)=1$	\\
     2	&$[0,6]$ &$x^2+1$	&$(0 \ 1 \ 0 \ $-$1 \dots)$	&$4$	&$\gcd(n,2)=1$	\\
     5	&$[$-$1,6]$&$x^2+x+1$	&$(0\ 1\ $-$1\dots)$	&$5$	&$\gcd(n,3)=1$	\\
     6	&$[$-$1,8]$&$x^2+1$	&$(0 \ 1 \ 0 \ $-$1\dots)$	&$6$	&$\gcd(n,2)=1$	\\
     7	&$[$-$1,8]$&$x^2+x+1$	&$(0\ 1\ $-$1\dots)$	&$6$	&$\gcd(n,3)=1$	\\
     10	&$[$-$1,9]$&$x^2+x+1$	&$(0\ 1\ $-$1\dots)$	&$6$	&$\gcd(n,3)=1$	\\
    \end{tabular}
\end{table}

Moreover, Boston raised the following question \cite{Boston_CyclicCodesAlgebraicGeometry_2001}:
\begin{question}[Boston's Question, \cite{Boston_CyclicCodesAlgebraicGeometry_2001}]
Let $3 \nmid n$ and the set $T = \{ 0,1,3,4,6,7,9,10,\dots,r\} \subseteq \defset$. Is the minimum distance $d$ then $d \geq \dBOSTON = |T|$?
\end{question}
Counter-examples show that Boston's conjecture is not true (see Example~\ref{ex:questionboston}), since the actual distance of such codes is not always $\dBOSTON = r+1$.
However, using the power series expansion
of $1/(x^2+x+1)$ with $\mathbf a = (0 \ 1 \ $-$1 \dots)$ we obtain
$\mu-1 = r+2$.
The minimum distance of such codes can be bounded by $d_f = \left\lceil (r+1)/2+1 \right\rceil$
with $u=\deg f(x) = 2$ and $v= h(x) =  1$.
\begin{example}[Distance of the $\CYC{\F{3}}{20}{6}{8}$ code]\label{ex:questionboston}
 Let $\defset = \{ 0,1,2,3,4,6,7,8,9,10,12,14,16,18\}$. For Boston's scheme, we can use
 $T = \{ 0,1,3,4,6,7,9,10,12\}$ with $|T| = 9$. The actual distance is $d=8$ and therefore, Boston's conjecture is not true.
 The BCH bound yields $\dBCH\geq6$. Our new bound is tight and with $r=12$, we obtain $d_f = \lceil (r+1)/2+1 \rceil = 8$.
\end{example}

\section{Generalized Key Equation and\\ Decoding Algorithm} \label{sec_decoding}
In this section, we present an efficient decoding algorithm up our new bound based
on a generalized key equation.

Let $(r_{0} \ r_{1}\ \dots \ r_{n-1})$ denote the received
word, i.e.,
$$
 (r_{0}\ r_{1}\ \dots \ r_{n-1})  = (c_{0}\ c_{1}\ \dots \ c_{n-1})+
 (e_{0}\
 e_{1}\  \dots \  e_{n-1}),
$$
and let $r(x) = \sum_{i=0}^{n-1} r_i x^i$ be the received polynomial.
Let $\mathcal E \subseteq \lbrace
0,\dots,n-1\rbrace$ be the set of error positions and let
$|\mathcal E|=t$. We define
the syndrome polynomial $S(x)$:
\begin{align}\label{eq:syndcalc}
S(x)  & \equiv  \sum \limits_{i=0}^{n-1} r_i \frac{\alpha^{ib} h(\alpha^{i}x)}{f(\alpha^{i}x)} \nonumber \\
&  = \sum \limits_{i \in \mathcal E}e_i \frac{\alpha^{ib} h(\alpha^{i}x)}{f(\alpha^{i}x)} \mod x^{\mu-1}.
\end{align}
Thus, the explicit form of the syndrome polynomial $S(x)$ is
\begin{equation}\label{eq:syndexplicit}
 S(x) = \sum \limits_{j=0}^{\mu-2} a_j r(\alpha^{j+b})x^j= \sum \limits_{j=0}^{\mu-2} a_j e(\alpha^{j+b})x^j.
\end{equation}
Based on the relation between the rational function $\alpha^{ib}\cdot h(\alpha^i x)/f(\alpha^ix)$ and
all codewords of a $q$-ary cyclic code \CYC{\Fq}{n}{k}{d} as defined in Definition~\ref{def_PowerSeriesCode}
in Section~\ref{sec_principle}, we introduce a generalized
error-locator polynomial $\Lambda(x)$ and error-evaluator polynomial $\Omega(x)$
and relate it to the syndrome definition of~\eqref{eq:syndcalc}.
Let $\mathcal{E}$ denote the set of error positions and let $t = |\mathcal{E}|$. We define $\Lambda(x)$ as:
\begin{equation} \label{eq_def_ELP}
 \Lambda(x) \defeq \prod_{i \in \mathcal E} f(\alpha^{i}x).
\end{equation}
Let
\begin{equation} \label{eq:omegaexplicit}
 \Omega(x) \defeq \sum \limits_{i \in \mathcal E} \Big(e_i \cdot\alpha^{ib}\cdot h(\alpha^{i}x) \cdot \prod_{\substack{j \in \mathcal E\\ j \neq i}} f(\alpha^jx) \Big),
\end{equation}
and we obtain with~\eqref{eq:syndcalc} a so-called generalized key equation:
\begin{equation} \label{eq:keyequation}
\begin{split}
 \Lambda(x)  \cdot S(x) & \equiv \Omega (x) \mod x^{\mu-1} \quad \text{with} \\
  \deg \Omega (x) & \leq (t-1)u+v \\
 & < \deg \Lambda(x) = tu,
\end{split}
\end{equation}
since $v < u$.

The main step of our decoding algorithm is to determine $\Lambda(x)$ and $\Omega(x)$ if $S(x)$ is given.
The following lemma shows that there is a unique solution for $\Lambda(x)$ if the number of errors is not too big.
\begin{lemma}[Solving the Key Equation]\label{lem:solveke}
 Let $S(x)$ with $\deg S(x) = \mu-2$ be given by \eqref{eq:syndexplicit}. If
 \begin{equation} \label{eq_numberoferrors}
  t = |\mathcal E|\leq \left\lfloor \frac{d_f-1}{2} \right\rfloor,
 \end{equation}
 there is a unique solution (up to a scalar factor) of the key equation \eqref{eq:keyequation} with $\deg \Omega (x) \leq (t-1)u+v < \deg \Lambda(x) = tu$. We can
 find this solution by the Extended Euclidean Algorithm (EEA) with the input polynomials $x^{\mu-1}$ and $S(x)$.
\end{lemma}
\begin{IEEEproof}
We use the properties of the EEA as proven in~\cite{Sugiyama_AMethodOfSolving_1975} (see also~\cite[Theorem~16, p. 367]{MacWilliamsSloane_TheTheoryOfErrorCorrecting_1988}). It guarantees the uniqueness (up to a scalar factor) of
the solution of~\refeq{eq:keyequation} and provides the stopping criteria of the EEA to obtain
$\Lambda(x)$ and $\Omega(x)$.

We require that $\deg \gcd(\Lambda(x), \Omega(x)) = 0$ (which follows from $\deg \gcd (f(x),h(x))=0$ and~\refeq{eq_def_ELP} and~\refeq{eq:omegaexplicit}).
Let the polynomials $x^{\mu-1}$ and $S(x)$ be given as input for the EEA and let the EEA stop as soon as the degree of the remainder $\deg r_i(x)$ in the $i$th step is less than or equal to  $\lfloor (\mu-1)/2 \rfloor$.
Then, we obtain the unique (except for a scalar factor) solution $\Lambda(x)$ and $\Omega(x)$ of~\eqref{eq:keyequation}, if~\eqref{eq_numberoferrors} holds.
For the explicit proof we refer to~\cite[Theorem~16, p. 367]{MacWilliamsSloane_TheTheoryOfErrorCorrecting_1988}. It shows that there is a unique solution of the generalized key equation~\eqref{eq:keyequation}
and that the EEA finds it if
\begin{equation}
\deg \Lambda(x) = tu \leq \left\lfloor\frac{\mu-1}{2}\right\rfloor,
\end{equation}
and therefore
\begin{equation} \label{eq_definitionOmega}
t \leq \left\lfloor\frac{\mu-1}{2u}\right\rfloor = \left\lfloor\frac{(d_f-1)u +v}{2u} \right\rfloor = \left\lfloor\frac{(d_f-1)}{2}\right\rfloor,
\end{equation}
since $v/2u < 1/2$.
\end{IEEEproof}
Key equation~\eqref{eq:keyequation} can be written as a linear system of equations, with $tu$ coefficients of a normalized $\Lambda(x)$ as unknowns.
If we consider only the equations which do not depend on $\Omega(x)$, we obtain:
 \begin{equation}\label{eq:keyeqsystem}
 \begin{pmatrix}
S_{tu}& S_{tu-1}& \dots& S_0\\
S_{tu+1}& S_{tu}& \dots& S_1\\
&&\vdots&\\
S_{\mu-2}& S_{\mu-3}& \dots& S_{\mu-tu-2}\\
 \end{pmatrix}
\cdot
\begin{pmatrix}
 1\\
 \Lambda_1\\
 \vdots\\
 \Lambda_{tu}
\end{pmatrix}
= \mathbf 0.
\end{equation}
There is a unique solution if and only if the rank of the syndrome matrix is $tu$.
One coefficient of $\Lambda(x)$ can be chosen arbitrarily (here $\Lambda_0=1$), since a scalar factor does not change the roots.
From this we obtain the same condition on the decoding radius as in Lemma~\ref{lem:solveke}.

If we have found $\Lambda(x)$, we can determine its factors $f(\alpha^i x)$, where $i\in \mathcal E$.
These factors are disjoint since $\deg(\gcd(f(\alpha^{i}x),f(\alpha^{j}x)))=0$, $\forall i
\neq j$ and therefore these factors provide the error positions.
We calculate only \emph{one} root $\beta_i$ of each $f(\alpha^ix)$ in a preprocessing step.
To find the error positions if $\Lambda(x)$ is given, we do a Chien search with $\beta_0,\beta_1,\dots,\beta_{n-1}$.
This is shown in Algorithm~\ref{algo:decalgo} and Theorem~\ref{theo:correctness} proves that each $\beta_i$ uniquely
determines $f(\alpha^i x)$.

For the non-binary case, we have to calculate the error values at the error positions.
This can be done by a generalized Forney's formula~\cite{Forney_OndecodingBCHcodes_1965}.
In order to obtain this error evaluation formula, we use the explicit expression for $\Omega(x)$ from~\eqref{eq:omegaexplicit}.
As mentioned before, the preprocessing step calculates $n$ values $\beta_0,\beta_1,\dots,\beta_{n-1}$ such that
\begin{equation*}
 f(\alpha^{i} \beta_{i}) = 0, \ \forall i = 0,\dots,n-1, \ \text{and} \ f(\alpha^{j}\beta_{i}) \neq 0, \ \forall j \neq i.
\end{equation*}
The evaluation of $\Omega(x)$ at $\beta_{\ell}$, $\ell \in \mathcal E$, yields:
\begin{equation*}
 \Omega(\beta_{\ell}) = \sum \limits_{i \in \mathcal E} \Big(e_i \cdot \alpha^{ib}\cdot h(\alpha^{i}\beta_{\ell}) \cdot \prod_{\substack{j \in \mathcal E\\ j \neq i}} f(\alpha^j \beta_{\ell}) \Big).
\end{equation*}
With $f(\alpha^{\ell} \beta_{\ell}) = 0$, the product $\prod_{{j \in \mathcal E, j \neq i}} f(\alpha^j \beta_{\ell})$ is zero if $\ell \in \mathcal E \backslash \lbrace i\rbrace$ and nonzero
only if $\ell = i$. Hence, we obtain
\begin{equation}\label{eq:omega-erroreval}
 \Omega(\beta_{\ell}) = e_{\ell}\cdot \alpha^{\ell b} \cdot h(\alpha^{\ell}\beta_{\ell}) \cdot \prod_{\substack{j \in \mathcal E\\ j \neq \ell}} f(\alpha^j\beta_{\ell}).
\end{equation}
This derivation provides the following lemma.
\begin{lemma}[Generalized Error Evaluation]\label{lem:errorevaluation}
Let the integer $b$, the polynomials $h(\alpha^{i}x)$, $f(\alpha^{i}x)$, $\Lambda(x) = \prod_{i \in \mathcal E} f(\alpha^{i}x)$ and $\Omega(x)$ from \eqref{eq:omegaexplicit}, for all $i=0,\dots,n-1$ with $\deg(\gcd(f(\alpha^{i}x),f(\alpha^{j}x)))=0$ be given.
 Then, the error values $e_{\ell}$ for all $\ell \in \mathcal E$ are given by
 \begin{equation}\label{eq:errorevaluation}
 \begin{split}
 e_{\ell} & = \frac{\Omega(\beta_{\ell})}{\alpha^{\ell b}\cdot h(\alpha^{\ell}\beta_{\ell})\prod_{\substack{j \in \mathcal E\\ j \neq \ell}} f(\alpha^j \beta_{\ell})} \\
 & = \frac{\Omega(\beta_{\ell}) \cdot f'(\alpha^{\ell}\beta_{\ell})}{\Lambda'(\beta_{\ell}) \cdot \alpha^{\ell b}\cdot h(\alpha^{\ell}\beta_{\ell})},
 \end{split}
\end{equation}
where $f'(\alpha^{i}x)$ and $\Lambda'(x)$ denote the derivatives of $f(\alpha^{i}x)$ and $\Lambda(x)$.
\end{lemma}
\begin{IEEEproof}
The lemma follows from~\eqref{eq:omega-erroreval} and the fact that
 \begin{equation*}
  \Lambda'(x) = \sum_{i \in \mathcal E} f'(\alpha^{i}x) \prod_{\substack{j \in \mathcal E\\ j \neq i}} f(\alpha^jx)
 \end{equation*}
and therefore
 \begin{equation*}
  \Lambda'(\beta_{\ell}) = f'(\alpha^{\ell}\beta_{\ell}) \prod_{\substack{j \in \mathcal E\\ j \neq \ell}} f(\alpha^j\beta_{\ell}).
 \end{equation*}
\end{IEEEproof}
Note that \eqref{eq:errorevaluation} is the classical Forney's formula~\cite{Forney_OndecodingBCHcodes_1965}, for
$f(\alpha^ix) = 1-\alpha^ix$ and $\alpha^{ib}\cdot h(\alpha^ix)=1$.

The decoding approach is summarized in Algorithm~\ref{algo:decalgo} and its correctness is proved in Theorem~\ref{theo:correctness}.

\printalgo{\caption{\texttt{Decoding $q$-ary Cyclic Codes}}
\label{algo:decalgo} \dontprintsemicolon \SetVline
\linesnumbered
\SetKwInput{KwPre}{\underline{Preprocessing}}
\SetKwInput{KwIn}{\underline{Input}}
\SetKwInput{KwOut}{\underline{Output}}
\BlankLine
\KwIn{Received word $r(x)$, $f(\alpha^{i}x)$,
$\alpha^{ib}\cdot h(\alpha^{i}x)$} 
\BlankLine 
\KwPre{Calculate one root of each
$f(\alpha^{i}x)$ $\Longrightarrow$ $\beta_0,\beta_1,\dots,\beta_{n-1}$}
\BlankLine 
Calculate $S(x)$ by~\eqref{eq:syndexplicit}\; 
\BlankLine
Solve Key Equation: Obtain $\Lambda(x)$, $\Omega(x)$ as output of \texttt{EEA($x^{\mu-1}, S(x)$)}\; 
\BlankLine
Chien--Search: Find all $i$ for
which $\Lambda(\beta_i)=0$, 
save them as $\widehat{\mathcal E}=\lbrace i_0,i_1,\dots,i_t\rbrace$\; 
\BlankLine
Error Evaluation: $\hat e_{\ell} = \Omega(\beta_{\ell})/\big(h(\alpha^{\ell}\beta_{\ell})\prod_{{j \in \mathcal E, j \neq \ell}} f(\alpha^j\beta_{\ell})\big)$, for all $\ell \in \widehat{\mathcal E}$\;
\BlankLine
$\widehat e(x)$ $\leftarrow$ $\sum_{\ell \in \widehat{\mathcal E}}\hat e_{\ell} x^{\ell}$ \;
\BlankLine
 $\widehat c(x)$ $\leftarrow$ $r(x)-\widehat e(x)$\;
\BlankLine \KwOut{Estimated codeword $\widehat c(x)$}\;}

\begin{theorem}[Correctness of Algorithm~\ref{algo:decalgo}] \label{theo:correctness}
If the distance $d(r(x),c(x)) \leq \left\lfloor (d_f-1)/2 \right\rfloor$ for some codeword $c(x) \in \mathcal C$, then Algorithm~\ref{algo:decalgo} returns $\hat c(x)=c(x)$ with complexity
$\mathcal O((\deg f(x)\cdot n)^2)$ operations.
\end{theorem}
\begin{IEEEproof}
Let $S(x)$ be defined by \eqref{eq:syndexplicit}.
 As shown in Lemma~\ref{lem:solveke}, we can then solve the key equation uniquely for $\Lambda(x)$ if $t\leq \left\lfloor (d_f-1)/2 \right\rfloor$. Therefore, we obtain
 $\Lambda(x) = \prod_{i \in \mathcal E} f(x,\alpha_{i})$ with $\deg \Lambda(x) = tu$ in Step~2 of Algorithm~\ref{algo:decalgo} and also $\Omega(x) \equiv \Lambda(x) \cdot S(x) \mod x^{\mu-1}$.
To explain the preprocessing and the Chien--search, we note that for each polynomial $a(x)$ of degree $u$ defined over $\F{q^s}$ there exists a splitting field, i.e., an extension field $\F{q^{us}}$ of $\F{q^{s}}$, in which $a(x)$ has $u$ roots.
Therefore, each $f(\alpha^ix)$ can be decomposed into $u = \deg f(\alpha^ix)$ linear factors over a field $\F{q^{us}}$.
These factors are disjoint since $\deg(\gcd(f(\alpha^{i}x),f(\alpha^{j}x)))=0$ and hence, \emph{one} root of $f(\alpha^ix)$ uniquely defines $f(\alpha^ix)$ and $i$.
Hence, $\Lambda(\beta_j)=0$ if and only if $j \in \mathcal E$ and Step~3 correctly identifies the error positions.

Lemma~\ref{lem:errorevaluation} proves the generalized error evaluation and therefore, if $d(r(x),c(x)) \leq \left\lfloor (d_f-1)/2 \right\rfloor$ for some codeword $c(x) \in \mathcal C$,  Algorithm~\ref{algo:decalgo} returns $\hat c(x)=c(x)$.

To prove the complexity, we note that the input polynomials $S(x)$ and $x^{\mu-1}$ of the EEA have degrees at most $\mu-2$ and $\mu-1$, respectively.
Therefore, the complexity of the EEA is quadratic in $\mu$, i.e., $\mathcal O(\mu^2) \approx \mathcal O((u \cdot d_{f})^2)$.
 The Chien--search and the generalized error evaluation require the same complexity as for the classical case, which is $\mathcal O(n^2)$.
Therefore, we can upper bound the complexity of Algorithm~\ref{algo:decalgo} by $\mathcal O((u\cdot n)^2)=\mathcal O((\deg f(x)\cdot n)^2)$.
\end{IEEEproof}

We consider the code from Example~\ref{ex_fraction}
to illustrate the decoding algorithm in the following.

\begin{example}[Decoding Binary Code]
We consider again the \CYC{\F{2}}{17}{9}{5} code and write explicitly the associated
power series $a(-4,\alpha^ix)$ in polynomial form:
\begin{equation}\label{eq:fraction17-9}
\begin{split}
a(-4,\alpha^ix) & =  \frac{\alpha^{i 13}\cdot h(\alpha^{i}x)}{f(\alpha^{i}x)} \\
& =  \frac{\alpha^{13i}+\alpha^{14i}x}{1+\alpha^i x + \alpha^{2i}
x^2} \\
& =  \alpha^{13i} +  \alpha^{15i} x^2 + \alpha^{16i} x^3 + \\
& \hspace{.4cm} \alpha^i x^5 + \alpha^{2i} x^6 + \alpha^{4i} x^8 \mod x^9.
\end{split}
\end{equation}
For the syndrome polynomial, we obtain with $\mu-1=9$ and \eqref{eq:syndcalc}, \eqref{eq:syndexplicit} and \eqref{eq:fraction17-9}:
\begin{align*}
 S(x) & = \sum \limits_{i=0}^{n-1} e_i \cdot (\alpha^{13i} + \alpha^{15i} x^2 + \dots + \alpha^{4i} x^8) \\
 & = \sum \limits_{i \in \mathcal E} (\alpha^{13i} + \alpha^{15i} x^2+  \dots + \alpha^{4i} x^8)\\
 & = r(\alpha^{13}) + r(\alpha^{15})x^2+ \dots + r(\alpha^4)x^8 \\
 & = S_{0} + S_{2} x^2+S_{3} x^3 + S_5 x^5 + S_6 x^6 + S_8 x^8.
\end{align*}
As in Algorithm~\ref{algo:decalgo}, we calculate \texttt{EEA} ($x^9,
S(x)$) and stop if the degree of the remainder is smaller than
$\lfloor (\mu-1)/2 \rfloor=4$. Assume, two errors occurred, then we
obtain $\Lambda(x)$ with $\deg \Lambda(x) = tu = 2\cdot 2 =4$.

Using the EEA is equivalent to solving the following system of equations for $\Lambda(x)$:
\begin{equation}
 \begin{pmatrix}
 0 & S_3 & S_2 & 0 & S_0\\
 S_5 & 0 & S_3 & S_2 & 0\\
 S_6 & S_5 & 0 & S_3 & S_2\\
 0 & S_6 & S_5 & 0 & S_3\\
 \end{pmatrix}
\cdot
\begin{pmatrix}
 1\\
 \Lambda_1\\
 \vdots\\
\Lambda_4
\end{pmatrix}
= \mathbf 0,
\end{equation}
and with both approaches,
$\Lambda(x)$ has the roots
$f(\alpha^{i}x)=(1+\alpha^i x + (\alpha^{i}x)^2)$, $\forall \
i \in \mathcal E$. We know that each
$f(\alpha^{i}x)=(1+\alpha^i x + (\alpha^{i}x)^2)$ has two
roots in $\F{2^8}$ which are unique. We have a look-up-table
with one root $\beta_i$ of each $f(\alpha_{i}x)$ and we do the
Chien search for $\Lambda(x)$ with $\beta_0,\beta_1,\dots,\beta_{n-1}$.
Since this is a binary code, we do not need an error evaluation and can
reconstruct the error.
\end{example}

\section{Conclusion} \label{sec_conclusion}
A new lower bound on the minimum distance of $q$-ary cyclic
codes is proved. For several classes of codes, a
more explicit bound on their distance is given. The connection
to existing bounds (BCH, HT and
Boston) is shown.

Furthermore, we derived a generalized key equation, which relates the syndrome definition and the polynomial for the determination of the error locations. This allows the realization of a quadratic-time decoding algorithm and provides an explicit expression for
the error evaluation.

\section{Acknowledgment}
The authors are grateful to  Maximilien Gadouleau for stimulating discussions.
We thank the anonymous referees for valuable comments that improved the presentation of this paper.


\newpage
\appendix
\begin{table}[htb]
\caption{Binary Codes and Bounds with $\mathbf a = (1 \ -1 \ 0)$ and $\mathbf a = (1 \ -1 \ 0 \ 0)$}
\label{tab:boundscodesq2}
\centering
\begin{tabular}{c|c|c|c|c}
$n$ & $\#$ codes & $\# \; \dBCH<d$ & $\# \; d_f > \dBCH$& $\# \; d_f <d$ \\
 \hline
15 & 32 & 2 & 2 &  0 \\
17*	&8	&2	&2	&0	\\
19	&4	&0	&0	&0	\\ 	
21	&8	&2	&2	&0	\\ 		
23	&8	&4	&0	&4	\\
25*	&8	&0	&0	&0 	\\
27	&0	&0	&0	&0 	\\
29	&4	&0	&0	&0	\\
31	&128	&34	&7	&31	\\
33*	&0	&0	&0	&0	\\
35	&64	&24	&8	&22	\\
37	&4	&0	&0	&0	\\
39	&0	&0	&0	&0	\\
41*	&8	&4	&4	&4	\\
43*	&16	&6	&3	&6	\\
45	&256 & 69	&22	& 57 \\
47	&8	&4	&0	&4	\\
49	&4	&0	&0	&0	\\
51	&256	&122	&4	&118	\\
53	&4	&0	&0	&0	\\
55	&32	&16	&4	&16	\\	
57*	&32	&10	&4	&10	\\
59	&4	&0	&0	&0	\\
61	&4	&0	&0	&0  \\
63	&8192	&4088	&509	&4088 \\
\end{tabular}
\end{table}
\begin{table}[htb]
\caption{Ternary Cyclic Codes and Bounds with $\mathbf a = (1 \ -1 \ 0)$ and $\mathbf a = (1 \ -1 \ 0 \ 0)$}
\label{tab:boundscodesq3}
\centering
\begin{tabular}{c|c|c|c|c}
$n$ & $\#$ codes & $\# \; \dBCH<d$ & $\# \; d_f > \dBCH$& $\# \; d_f <d$\\
 \hline
8	&32	&2	&2	&0	\\
11	&8	&4	&2	&4	\\
13	&32	&6	&0	&0	\\
16	&128	&16	&8	&8	\\
20	&128	&38	&6	&36	\\
22	&64	&40	&22	&40	\\
23	&8	&4	&0	&4	\\
26	&1024	&512	&108	&490	\\
28	&128	&18	&2	&18	\\
32	&512	&102	&46	&57	\\
35	&32	&16	&2	&16	\\
37	&8	&4	&0	&4	\\
\end{tabular}
\end{table}



\newpage

\textbf{Alexander Zeh} studied electrical engineering at the University of Applied Science in Stuttgart, with the main topic automation technology.
He received his Dipl.-Ing. (BA) degree in 2004. He continued his studies at Universität Stuttgart until 2008, where he received
is Dipl.-Ing. in electrical engineering. He participated in the double-diploma program with Télécom ParisTech (former ENST) from
2006 to 2008 and he received also a french diploma.
Currently he is a Ph.D. student at the Institute of Communications Engineering, University of Ulm, Germany
and at the Computer Science Department (LIX), École Polytechnique ParisTech, Paris, France.
His current research interests include coding and information theory, signal processing, telecommunications and the implementation of fast algorithms on FPGAs.
\vspace{1ex}

\textbf{Antonia Wachter-Zeh} 
studied electrical engineering at the University of Applied Science in Ravensburg, with the main topic communication technology.
She received her Dipl.-Ing. (BA) degree in 2007.
She continued her studies at the University of Ulm until 2009, where she received her M.Sc. in electrical engineering.
She is currently working towards the Ph.D. degree at the Institute of Communications Engineering, University of Ulm, Germany 
and at the Institut de recherche mathématique de Rennes (IRMAR), Université de Rennes 1, Rennes, France.
Her major research interests are topics in coding theory.

\vspace{1ex}

\textbf{Sergey V. Bezzateev} was born in Leningrad, Soviet
Union, on June 10, 1957. He received his diploma in computer
science from the Airspace Instrumentation Institute of Leningrad,
Soviet Union in 1980. In 1987, he received his Ph.D. degree in
information theory from the Airspace Instrumentation Institute
of Leningrad. From 1980 to 1993, he was employed by the Airspace
Instrumentation Institute. From 1993 to 1995 he worked as researcher at
the Nagoya University, Japan, where he co-operated with Prof.
Yoshihiro Iwadare. In 1995, he became associate professor at the Department of Information Technologies and
Information Security, State University of Airspace
Instrumentation (SUAI), Saint Petersburg, Russia. From 2004
till 2007, he was project leader of the Joint Laboratory
Samsung-SUAI on Information Security in Wireless Networks.
In 2010, he became Professor and the head of Department of 
Technologies of Information Security in SUAI.
His main research interests include coding theory and cryptography.

\vspace{1ex}

\end{document}